\documentclass[12pt]{article}
\usepackage{graphicx}
\usepackage{amsfonts,amssymb}
\usepackage{amsmath}
\usepackage{citehack,cite}

\newcommand{\non}{\nonumber \\}

\newcommand{\be}{\begin{equation}}
\newcommand{\ee}{\end{equation}}
\newcommand{\bea}{\begin{eqnarray}}
\newcommand{\eea}{\end{eqnarray}}

\newcommand{\lp}{\left (}
\newcommand{\rp}{\right )}
\newcommand{\lb}{\left [}
\newcommand{\rb}{\right ]}
\newcommand{\lbr}{\left \{}
\newcommand{\rbr}{\right \}}
\newcommand{\ld}{\left .}
\newcommand{\rd}{\right .}

\begin{document}

\begin{center}
{\bf FIRST-ORDER PHASE TRANSITION IN THE FRAMEWORK OF THE CELL FLUID
MODEL: REGIONS OF CHEMICAL POTENTIAL VARIATION AND THE CORRESPONDING
DENSITIES}
\end{center}

\begin{center}
{\sc I.V. Pylyuk\footnote{e-mail: piv@icmp.lviv.ua}, M.P. Kozlovskii}
\end{center}

\begin{center}
{\it Institute for Condensed Matter Physics  \\
of the National Academy of Sciences of Ukraine, \\
1~Svientsitskii Str., 79011 Lviv, Ukraine}
\end{center}

\vspace{0.5cm}

{\small
A microscopic description is given for the behavior of the fluid system in an immediate vicinity
of its critical point, where theoretical and experimental researches are difficult to carry out.
For the temperatures $T<T_c$, the regions of chemical potential and density variations are singled
out and analyzed. The equation of state of the cell fluid model in terms of temperature-chemical
potential is written using the Heaviside functions. This equation is also given in terms of the
temperature and density variables. As a result of the study of the relationship between the
density and the chemical potential, an equation for the binodal curve is obtained in a narrow
neighborhood of the critical point.
}

\vspace{0.5cm}

PACS numbers: 05.70.Ce, 64.60.F-, 64.70.F-

Keywords: cell fluid model, chemical potential, density, equation of state, binodal

\section{Introduction}
\label{sec:1}

The theoretical and experimental study of the behavior of liquids and
their mixtures in a vicinity of their critical point (see, e.g.,
works \cite{l104,b112,p112,y114,y118,v118,o197,p104}) is an important
and challenging task. In our previous works \cite{kpd118,p120},
the behavior of fluid was studied in an immediate
vicinity of the critical point, and in works \cite{kd116,kdp117,pd120}
beyond this vicinity. As a result, a wide region near
the critical point has been covered. The mathematical
description was carried out in the framework of
the cell fluid model using the grand canonical ensemble.
The whole volume $V$ of a system consisting
of $N$ interacting particles was conventionally divided
into $N_v$ cells, each of the volume $v=V/N_v=c^3$,
where $c$ is the linear cell size. Note that unlike the
lattice gas model where the cell is assumed to contain
no more than one particle, in this approach the
cell can include more than one particle.

In works \cite{kpd118,p120}, the analysis was performed in the
framework of the collective-variables approach
using the renormalization group transformation \cite{ymo287}.
An analytical procedure for calculating the grand
partition function and the thermodynamic potential
of the cell fluid model was developed in works
\cite{kpd118,p120} in the approximation of a non-Gaussian
(quartic) distribution for order parameter fluctuations
without involving the hard-spheres reference
system. The formation of the reference system
as a part of the repulsive component of the interaction
potential made it possible to take into account all
kinds of interaction (both short- and long-range) from
the same position of the collective-variables approach.

The role of interaction potential in this work is
played by the Morse potential. The interaction potential
parameters, which are given in works \cite{kpd118,p120} and
are necessary for quantitative estimates, correspond
to the data for sodium. They were taken from work
\cite{singh}, which was devoted to the study of vapor-liquid
equilibrium curves for metals using Monte Carlo simulation
and Morse potential. Taking this potential
into account, the vapor-liquid coexistence curves for
metals were also analyzed in the framework of the
integral-equations approach \cite{apf_11}.

The Morse potential is widely used when studying
the melting and laser ablation processes using computer
simulation \cite{z102,x104,l102}. There are works in the literature
devoted to the study of the structural properties
of the Morse and Lennard-Jones clusters, as well
as to a comparison between them \cite{l102,d196,d104,t102}. In some
cases, modifications of the Morse function were made
\cite{c198,s199,l105,d198} in order to improve the numerical results.
Although the Morse potential was traditionally used to
model covalently bound diatomic molecules \cite{m129,l199,c105}, it
is also applied to estimate non-bounding interactions
\cite{m101,o103}. This potential is qualitatively similar to the
Lennard-Jones potential, but they are quite different
from the quantitative point of view. The Lennard-
Jones and Morse potentials can be compared directly
using a mathematical relationship enabling the point
of energy minimum to be located at the same position
\cite{s102,okumura_00}. In addition, it was shown that either of the
potentials can be derived from the other \cite{l103}.

This work complements the study of the critical
behavior of the Morse fluid that was performed in
works \cite{kpd118,p120}. In particular, for temperatures lower
than the critical one, solutions of a certain cubic equation
are obtained. They govern the quantities entering
the equation of state of the cell fluid model. The
behavior of the equation solutions depending on the
chemical potential value is analyzed in the immediate
vicinity of the critical point, and the regions where
the chemical potential changes are singled out. Each
of those regions is considered separately. Expressions
for the boundary densities (the densities at the regions'
boundaries) are obtained. For this purpose, a
nonlinear equation, which relates the density to the
chemical potential, is used. The equation of state of
the cell fluid model and the binodal equation are presented.

\section{Chemical potential and density changes
at temperatures below the critical one}
\label{sec:2}

The equation of state of the cell fluid model for temperatures
$T<T_c$ \cite{p120} contains quantities dependent
on the solution $\sigma'_0$ of the equation
\be
(\sigma'_0)^3 + p' \sigma'_0 + q' = 0.
\label{3d12fb}
\ee
(the solution is also given in work \cite{p120}). The coefficients
\[
p' = 6 \frac{r_{n'_p+2}}{u_{n'_p+2}}, \quad
q' = - 6 \frac{s^{5/2}}{u_{n'_p+2}} \frac{\tilde h}{(\tilde h^2 + h^2_{cm})^{1/2}}
\]
include the quantities $r_{n'_p+2}$ and $u_{n'_p+2}$ determining
the long-wavelength part of the grand partition function
of the model. The quantity $\tilde h$ is proportional to
the chemical potential $M$, and the quantity $h_{cm}$ is
characterized by the renormalized relative temperature
$\tau = (T-T_c)/T_c$ (see work \cite{p120}). The renormalization
group parameter $s$ determines the separation
of the phase space of collective variables into layers.

The form of the solutions of Eq. (\ref{3d12fb}) depends on the
sign of the discriminant
\be
Q = (p'/3)^3 + (q'/2)^2.
\label{3d13fb}
\ee
If $Q>0$, the single real solution $\sigma'_0$ of Eq. (\ref{3d12fb}), according
to Cardano's formula, looks like
\bea
&&
\sigma'_{0b} = A+B, \non
&&
A = (-q'/2 + Q^{1/2})^{1/3}, \non
&&
B = (-q' / 2 - Q^{1/2})^{1/3}.
\label{3d15fb}
\eea
If $Q<0$, there are three real solutions (the quantity
$\sigma'_0$ acquires three possible real values)
\bea
&&
\sigma'_{01} = 2 (-p' / 3)^{1/2} \cos (\alpha_r / 3), \non
&&
\sigma'_{02,03} = -2 (-p' / 3)^{1/2} \cos (\alpha_r / 3 \pm \pi / 3),
\label{3d16fb}
\eea
where $\alpha_r$ is determined from the equation
\be
\cos \alpha_r = - \frac{q'}{2(-p'/3)^{3/2}}.
\label{3d17fb}
\ee

If the discriminant is negative, solutions (\ref{3d16fb}) can be
rewritten as follows:
\bea
&&
\sigma'_{01} = 2\sigma_{0r} \cos\frac{\alpha_r}{3}, \quad
\sigma'_{02} = - 2\sigma_{0r} \cos\left( \frac{\alpha_r}{3} + \frac{\pi}{3}\right),\non
&&
\sigma'_{03} = - 2\sigma_{0r} \cos\left( \frac{\alpha_r}{3} - \frac{\pi}{3}\right).
\label{5d1fb}
\eea
Here,
\bea
&&
\sigma_{0r} = \left( - \frac{2r_{n'_p+2}}{u_{n'_p+2}}\right)^{1/2}, \non
&&
\alpha_r = \arccos \left( \frac{M\left( \tilde h^2_q+h^2_{cm}\right)^{1/2}}
{M_q\left( \tilde h^2+h^2_{cm}\right)^{1/2}}\right).
\label{5d2fb}
\eea
The chemical potential $M_q$ is determined from the condition $Q=0$ and
satisfies the equality
\be
M_q = \Biggl[ - \frac{8 r^3_{n'_p+2} (1 + \alpha^2_{mq})}{9 u_{n'_p+2} s^5 \beta W(0)}
\Biggr]^{1/2} h_{cm},
\label{3d14fb}
\ee
where
\be
\alpha_{mq} = \tilde h_q/h_{cm}, \quad
\tilde h_q = M_q \left( \beta W(0)\right)^{1/2}.
\label{5d3fb}
\ee
Here, $\beta = 1/(kT)$ is the inverse temperature, and
$W(0)$ is the Fourier transform of the effective interaction
potential \cite{kpd118} at the zero wave vector. For all
$|M|<M_q$, the discriminant $Q<0$ and, therefore,
there are three real roots of Eq. (\ref{3d12fb}) in this interval of
$M$-values.

\begin{figure}
\centering \includegraphics[width=0.70\textwidth]{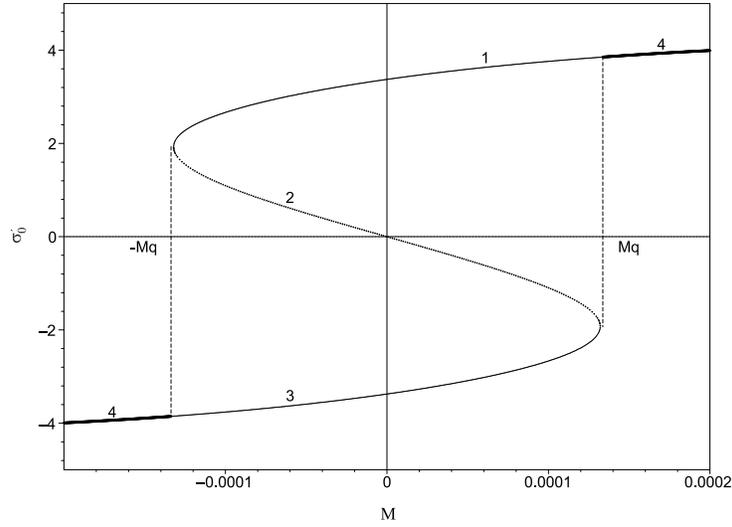}
\caption{Solutions of cubic equation (\ref{3d12fb}) as functions of the
chemical potential $M$ at $\tau=-0.005$. Curves 1, 2, 3, and 4 correspond
to $\sigma'_0=\sigma'_{01}$, $\sigma'_0=\sigma'_{02}$,
$\sigma'_0=\sigma'_{03}$, and $\sigma'_0=\sigma'_{0b}$, respectively.}
\label{fig_1fb}
\end{figure}

\begin{figure}
\centering \includegraphics[width=0.70\textwidth]{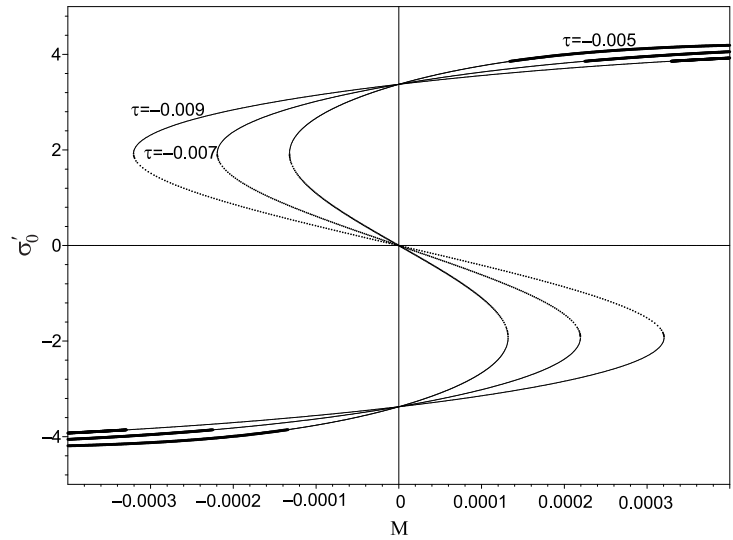}
\caption{Dependence of the solutions of cubic equation (\ref{3d12fb})
on the chemical potential $M$ at various values of the relative temperature
($\tau=-0.005$, $\tau=-0.007$, $\tau=-0.009$).}
\label{fig_2fb}
\end{figure}

The dependence of the solutions of the cubic equation
(\ref{3d12fb}) on the chemical potential $M$ at $T<T_c$ is
shown in Figs.~\ref{fig_1fb} and \ref{fig_2fb}.
Figure~\ref{fig_2fb} makes it possible to trace the shift of $M_q$
(the joint points of the thin and thick solid curves) as $\tau$ changes.
As one can see, the absolute value of $M_q$ decreases with the reduction
of $|\tau|$.

In the case $|M|>M_q$, as was at $T>T_c$, Eq. (\ref{3d12fb}) has
a single real solution (\ref{3d15fb}). For the latter, at $M=-M_q$
and taking into account the equality $Q=0$, we obtain
\be
\sigma_{0bq}^{'(-)} \! = \! 2 \left[ - 3s^{5/2} \frac{M_q(\beta W(0))^{1/2}}
{u_{n'_p+2} h_{cm}(1+\alpha^2_{mq})^{1/2}}\right]^{1/3} \!\!\! = \!\! - 2\sigma_{0r},
\label{5d4fb}
\ee
whereas at $M=M_q$ we have
\be
\sigma_{0bq}^{'(+)} = 2 \left[  3s^{5/2} \frac{M_q(\beta W(0))^{1/2}}
{u_{n'_p+2} h_{cm}(1+\alpha^2_{mq})^{1/2}}\right]^{1/3} = 2\sigma_{0r}.
\label{5d5fb}
\ee

Let us analyze the asymptotics of solutions (\ref{5d1fb}) at
$|M|=M_q$. If $M=-M_q$, we obtain $\cos \alpha_{rq}^{(-)}=-1$,
$\alpha_{rq}^{(-)}=\pi$, and
\bea
&&
\sigma_{01}^{'(-)} = 2 \sigma_{0r} \cos\frac{\pi}{3} = \sigma_{0r},\non
&&
\sigma_{02}^{'(-)} = - 2 \sigma_{0r} \cos\left( \frac{2\pi}{3}\right) = \sigma_{0r},\non
&&
\sigma_{03}^{'(-)} = - 2 \sigma_{0r} \cos 0 = - 2\sigma_{0r}.
\label{5d6fb}
\eea
The case $M=M_q$ brings us to the formulas $\cos \alpha_{rq}^{(+)}=1$,
$\alpha_{rq}^{(+)}=0$, and
\bea
&&
\sigma_{01}^{'(+)} = 2 \sigma_{0r} \cos 0 = 2 \sigma_{0r},\non
&&
\sigma_{02}^{'(+)} = - 2 \sigma_{0r} \cos \frac{\pi}{3} = - \sigma_{0r},\non
&&
\sigma_{03}^{'(+)} = - 2 \sigma_{0r} \cos \left( - \frac{\pi}{3} \right) = - \sigma_{0r}.
\label{5d7fb}
\eea
Thus, if $M=-M_q$, the solution is $\sigma_{0bq}^{'(-)}=-2\sigma_{0r}$ (see
Eq. (\ref{5d4fb})), which coincides with $\sigma_{03}^{'(-)}$ from
Eq. (\ref{5d6fb}). If $M=M_q$, the solution is
$\sigma_{0bq}^{'(+)}=2\sigma_{0r}$ (see Eq. (\ref{5d5fb}))
and it coincides with $\sigma_{01}^{'(+)}$ from Eq. (\ref{5d7fb}).

\begin{figure}
\centering \includegraphics[width=0.70\textwidth]{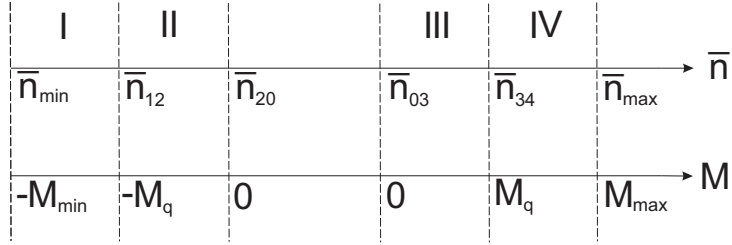}
\caption{Regions of chemical potential variation and the corresponding
densities for temperatures below the critical one.}
\label{fig_3fb}
\end{figure}

The conclusion drawn from the above calculations
is as follows. As the chemical potential increases to
$-M_q$ from the side of negative values, Eq. (\ref{3d12fb}) has a
single solution given by Eq. (\ref{3d15fb}) (region I, gas phase,
in Fig.~\ref{fig_3fb}). At $-M_q<M<0$, it transforms into the solution
$\sigma'_{03}$ from Eq. (\ref{5d1fb}), which is valid up to $M=-0$
(region II, transient gas phase). For $M=-M_q$, we obtain
$\sigma'_{03}=\sigma_{03}^{'(-)}=-2\sigma_{0r}$ (see Eq. (\ref{5d6fb})),
and for $M=-0$, we arrive at the expressions $\cos \alpha_{r0}^{(-)}=0$,
$\alpha_{r0}^{(-)}=\pi/2$, and
\be
\lim_{M\rightarrow-0}\sigma'_{03} = \sigma_{030}^{'(-)} =
- 2 \sigma_{0r} \cos\lp - \frac{\pi}{6}\rp = - \sqrt 3 \sigma_{0r}.
\label{5d8fb}
\ee
On the other hand, as $M$ decreases to $M_q$ from
the side of positive values, there is the single solution
(\ref{3d15fb}) (region IV, fluid phase). At $0<M<M_q$, this
solution transforms into the solution $\sigma'_{01}$ from
Eq. (\ref{5d1fb}), which is valid up to $M=+0$ (region III, transient
fluid phase). For $M=M_q$, we have
$\sigma'_{01}=\sigma_{01}^{'(+)}=2\sigma_{0r}$ (see Eq. (\ref{5d7fb})),
and for $M=+0$ we obtain
\be
\lim_{M\rightarrow +0}\sigma'_{01} = \sigma_{010}^{'(+)} =
2 \sigma_{0r} \cos\lp \frac{\pi}{6}\rp = \sqrt 3 \sigma_{0r}.
\label{5d9fb}
\ee

According to the chemical potential value $M$, the equation of state of
the cell fluid model at $T<T_c$ (see work \cite{p120}) can be written
in the form
\bea
&&
\frac{Pv}{kT} = P_a^{(-)}(T) + E_{\mu} +
D_{13} (\sigma'_{0b}) \left[ \Theta (-M-M_q) + \rd \non
&&
\ld + \Theta (M - M_q)\right] +
D_{13} (\sigma'_{03}) \Theta(-M) \Theta (M+M_q) + \non
&&
+ D_{13} (\sigma'_{01}) \Theta(M) \Theta(M_q-M).
\label{5d10fb}
\eea
Here, the quantity
\bea
&&
D_{13}(\sigma'_0) = \lp \gamma_s^{(-)} - e_2^{(-)}\rp
\lp \tilde h^2 + h^2_{cm}\rp^{\frac{d}{d+2}} + \non
&&
+ e_0^{(-)} \tilde h \lp \tilde h^2 + h^2_{cm}\rp^{\frac{d-2}{2(d+2)}}
\label{5d11fb}
\eea
depends on the solution $\sigma'_0$ of Eq. (\ref{3d12fb}); $d=3$ is
the space dimension; $v$ is the cell volume; $\Theta(M)$ is
the Heaviside function, which is equal to unity if $M>0$,
to zero if $M<0$, and to $1/2$ if $M=0$; the quantity $P_a^{(-)}(T)$
depends analytically on the temperature; the coefficient $\gamma_s^{(-)}$
characterizes the non-analytical contribution to the thermodynamic
potential; the quantities $e_0^{(-)}$ and $e_2^{(-)}$ depend on the roots
of the cubic equation (\ref{3d12fb}). Expressions for all those
quantities, as well as for $E_{\mu}$, are given in work \cite{p120}.

The equation of state (\ref{5d10fb}) makes it possible to study
the dependence of the pressure $P$ on the chemical potential $M$ and
the relative temperature $\tau$. This equation can be rewritten in terms
of the temperature and density variables. For this purpose, the chemical
potential expressed via the temperature $\tau$ and the average
density $\bar n$ should be substituted into Eq. (\ref{5d10fb}),
and the intervals of chemical potential values in the
Heaviside functions have to be replaced by the corresponding
density values from the regions shown in Fig.~\ref{fig_3fb}.
Let us consider each of the regions separately.

{\bf Region I ($M\leq -M_q$).} Here the solution $\sigma'_0$ of
Eq. (\ref{3d12fb}) looks like $\sigma'_{0b}$ from Eq. (\ref{3d15fb}).
If $M=-M_q$, then expression (\ref{5d4fb}), where $\sigma_{0r}$ is given
by the relationship from Eq. (\ref{5d2fb}), is valid for $\sigma'_{0b}$.
Solution (\ref{5d4fb}) for $\sigma_{0bq}^{'(-)}$ coincides with
$\sigma_{03}^{'(-)}$ from Eq. (\ref{5d6fb}). On the other hand,
the equality \cite{p120}
\be
b_3^{(-)} M^{1/5}  = \bar n - n_g + M,
\label{4d37fb}
\ee
holds, which couples the average density $\bar n$ with the
chemical potential (in this case, $M=-M_q$) and the
quantity $\sigma_{00}^{(-)}=f(\sigma'_{0b})$, which is included
into the coefficient $b_3^{(-)}$. Note that $n_g$ is determined via
the coefficients in the initial expression for the grand partition
function, and $\sigma_{00}^{(-)}$ is a function of the quantity
$\alpha_m = \tilde h/h_{cm}$, which includes the initial chemical
potential $\mu$ (included into $M$) and the relative
temperature $\tau$. From Eq. (\ref{4d37fb}), we can determine
the density $\bar n_{12}$ (the boundary density between
regions I and II) corresponding to the value
$M=-M_q$. Neglecting the last term in Eq. (\ref{4d37fb}), we
obtain
\bea
&&
\bar n_{12} = n_g + b_3^{(-)} M^{1/5}\big|_{\substack{M=-M_q}} =
n_g + \left[ (1+\alpha^2_{mq})^{1/2}h_{cm}\right]^{1/5}
\sigma_{00}^{(-)}(\sigma_{03}^{'(-)}), \non
&&
\sigma_{03}^{'(-)} = - 2 \sigma_{0r}.
\label{5d12fb}
\eea

{\bf Region II ($-M_q < M \leq -0$).} At $M=-0$, the equality
(\ref{5d8fb}) holds true and the boundary density
$\bar n_{20}=\lim_{M\rightarrow -0} \bar n$ takes the form
\bea
&&
\bar n_{20} = n_g +
\lim_{M\rightarrow -0} \left[ (1+\alpha^2_{m})^{1/2}h_{cm}\right]^{1/5}
\sigma_{00}^{(-)} =
n_g + h_{cm}^{1/5} \sigma_{00}^{(-)}(\sigma_{030}^{'(-)}), \non
&&
\sigma_{030}^{'(-)} = - \sqrt 3 \sigma_{0r}.
\label{5d13fb}
\eea

{\bf Region III ($+0 \leq M < M_q$).} This region starts from the value
$M=+0$, where $\sigma_{010}^{'(+)}=\sqrt 3 \sigma_{0r}$ and, accordingly,
\be
\bar n_{03} = n_g + h_{cm}^{1/5} \sigma_{00}^{(-)} (\sigma_{010}^{'(+)}).
\label{5d14fb}
\ee
The chemical potential $M$ in region III acquires values less than $M_q$.

{\bf Region IV ($M \geq M_q$).} This region starts from the value
$M=M_q$, which corresponds to $\sigma_{01}^{'(+)}=2\sigma_{0r}$,
so that
\be
\bar n_{34} = n_g + \left[ (1 + \alpha^2_{mq})^{1/2} h_{cm}\right]^{1/5}
\sigma_{00}^{(-)} (\sigma_{01}^{'(+)}).
\label{5d15fb}
\ee
The initial boundary density $\bar n_{34}$ in region IV increases
to a certain value $\bar n_{max}$, which corresponds to $M_{max}$.
At $\bar n > \bar n_{max}$, the chemical potential $M$ decreases with
the increasing density $\bar n$, which does not reflect the physical
nature of the phenomenon (a similar picture is observed
at $\bar n < \bar n_{min}$).

\begin{figure}
\centering \includegraphics[width=0.70\textwidth]{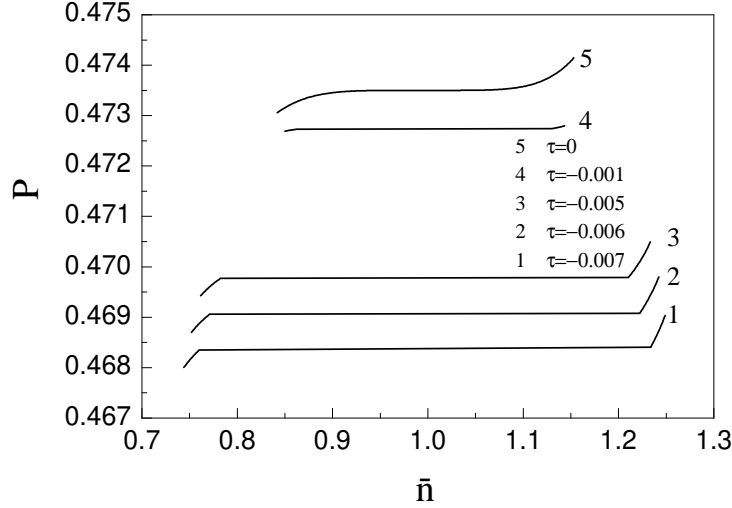}
\caption{Pressure as a function of average density for various values
of the relative temperature.}
\label{fig_4fb}
\end{figure}

The determination of the boundary densities $\bar n_{12}$, $\bar n_{20}$,
$\bar n_{03}$, and $\bar n_{34}$ makes it possible to write the equation
of state (\ref{5d10fb}) in the form
\bea
&&
\frac{Pv}{kT} = P_a^{(-)}(T) + E_{\mu} +
D_{13} (\sigma'_{0b}) \left[ \Theta (\bar n_{12}-\bar n) + \rd \non
&&
\ld + \Theta (\bar n - \bar n_{34})\right] +
D_{13} (\sigma'_{03}) \Theta(\bar n - \bar n_{12}) \Theta (-\bar n + \bar n_{20}) + \non
&&
+ D_{13} (\sigma'_{01}) \Theta(\bar n - \bar n_{03}) \Theta(\bar n_{34}-\bar n),
\label{5d16fb}
\eea
where
\be
D_{13}(\sigma'_0) \! = \! \lp \frac{\bar n \! - \! n_g}{\sigma_{00}^{(-)}} \rp^6
\!\! \left[ \! e_0^{(-)} \! \frac{\alpha_m}{(1+\alpha_m^2)^{1/2}} \! + \!
\gamma_s^{(-)} \! - \! e_2^{(-)} \! \right].
\label{5d17fb}
\ee
The dependence of the pressure $P$ (see Eq. (\ref{5d16fb})) on $\bar n$
for various $\tau$ is shown in Fig.~\ref{fig_4fb}.

\section{Relationship between the density
and the chemical potential of fluid.
Limiting cases}
\label{sec:3}

Nonlinear equation (\ref{4d37fb}), which describes the relationship
between the density $\bar n$ and the chemical potential $M$, can be
rewritten in the form \cite{p120}
\be
\bar n = n_g - M + \sigma_{00}^{(-)} \lp \tilde h^2 + h^2_{cm} \rp^{\frac{d-2}{2(d+2)}}.
\label{6d1fb}
\ee
The general form of equation (\ref{6d1fb}) or (\ref{4d37fb}) makes it
possible to change in a natural way to the cases when
either of the variables (the temperature or the chemical
potential) is crucial for the description of the order
parameter behavior.

Let us describe the behavior of $\bar n$ for some limiting
cases. One of them is the absence of chemical potential $M$ (i.e., $M=0$
and hence $\tilde h=0$) and $T\neq T_c$. Then, we obtain
\be
\sigma_{00}^{(-)}(M=0) = \frac{e_0^{(-)}}{(\beta W(0))^{1/2}} + e_{020}^{(-)},
\label{6d2fb}
\ee
where
\[
e_{020}^{(-)} = e_{02}^{(-)}(M=0) = \frac{1}{(\beta W(0))^{1/2}}
\frac{f_{Iv}}{s^3}.
\]
The expression for $f_{Iv}$ is given in work \cite{p120}. From
Eq. (\ref{6d1fb}), we obtain the dependence
\be
\bar n = n_g + \sigma_{00}^{(-)}(M=0) \tilde\tau_1^{\beta},
\label{6d3fb}
\ee
where the critical exponent $\beta=\nu/2$.

Another limiting case is $M\neq 0$ and $T=T_c$. The density $\bar n$ in
Eq. (\ref{6d1fb}) at $T=T_c$ satisfies the equality
\be
\bar n = n_g - M + \sigma_{00}^{(-)}(T_c) \tilde h^{1/\delta},
\label{6d4fb}
\ee
where
\be
\sigma_{00}^{(-)}(T_c) = \frac{6}{5} \frac{1}{(\beta_c W(0))^{1/2}}
\lb e_0^{(-)} + \gamma_s^{(-)} - e_2^{(-)} \rb
\label{6d5fb}
\ee
and the critical exponent $\delta=5$.

In the general case, i.e., at $M\neq 0$ and $T\neq T_c$,
Eq. (\ref{6d1fb}) can be presented as follows:
\be
\bar n = n_g - M + \sigma_{00}^{(-)} \lp \tilde h^2 +
\tilde\tau_1^{2\beta\delta} \rp^{1/(2\delta)}.
\label{6d6fb}
\ee
Note that $M\ll 1$, and $\tilde h\sim M$. Therefore, the second
summand, $M$, in the right-hand sides of Eqs. (\ref{6d1fb}),
(\ref{6d4fb}), and (\ref{6d6fb}) is much smaller than the third term
and can be neglected.

\section{Binodal equation}
\label{sec:4}

The binodal equation can be obtained from Eq. (\ref{6d1fb}) by
putting $M=0$. Then we arrive at Eq. (\ref{6d3fb}). Now, substituting
the expression $\tilde\tau_1 = -\tau \frac{c_{11}}{q} E_2^{n_0}$, we
obtain
\be
\bar n = n_g + \sigma_{00}^{(-)}(M=0)
\lp -\tau \frac{c_{11}}{q} E_2^{n_0} \rp^{\beta}.
\label{7d1fb}
\ee
Here, $E_2$ is one of the eigenvalues of the matrix for the
linear transformation of the renormalization group,
the quantity $c_{11}$ characterizes one of the coefficients in
the solutions of recurrence relations for the $\rho^4$-model
\cite{p120,kpd118}, $n_0$ is the difference between the exit points
from the critical fluctuation regime at $T>T_c$ and $T<T_c$, and $q$ is
associated with the averaging of the wave vector square. For more
information on those parameters, see work \cite{p120}.

\begin{figure}
\centering \includegraphics[width=0.70\textwidth]{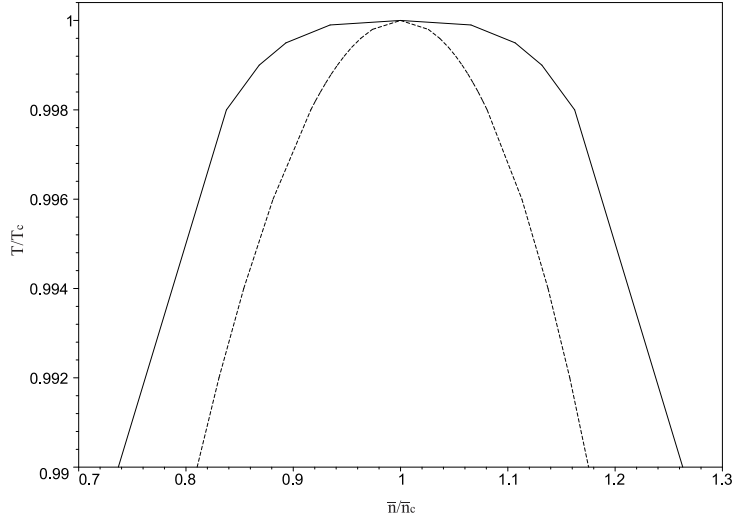}
\caption{Coexistence curve (binodal curve) obtained in an
immediate vicinity of the critical point taking into account
the interaction potential parameters that are characteristic of
sodium. The solid curve (the dome) was plotted according to
the obtained binodal equation, and the dotted curve is a result
of the zero-mode approximation \cite{kdp117}.}
\label{fig_5fb}
\end{figure}

Let us solve Eq. (\ref{7d1fb}) with respect to the temperature.
Taking the equalities $\tau=T/T_c-1$ and $\beta=\nu/2$ into account,
we obtain
\be
\lb \frac{\lp \frac{\bar n}{n_g}-1 \rp n_g}{\sigma_{00}^{(-)}(M=0)} \rb^{2/\nu}
\frac{q}{c_{11} E_2^{n_0}} = - \frac{T}{T_c} + 1
\label{7d2fb}
\ee
or
\be
\frac{T}{T_c} = 1 - \lbr \lb \frac{\lp \frac{\bar n}{n_g}-1 \rp n_g}
{\sigma_{00}^{(-)}(M=0)} \rb^2 \rbr^{1/\nu} \frac{q}{c_{11} E_2^{n_0}}.
\label{7d3fb}
\ee
On the basis of Eq. (\ref{7d3fb}), we can plot the binodal
curve in the temperature versus density plane (see Fig.~\ref{fig_5fb}).
This curve agrees with the data predicted for
sodium by extrapolating the results of computer simulation
\cite{singh} to $T/T_c \approx 1$ (see work \cite{p120}).

The spinodal equation, which describes the limiting
states of the system that determine the boundaries
of the instability region, can be found from the
extremum condition
\[
\frac{\partial (P v/k T)}{\partial \bar n}\bigg|_T= 0
\]
for the equation of state (\ref{5d16fb}), where we should
substitute the chemical potential $M$ expressed from
Eq. (\ref{4d37fb}) in terms of the average density $\bar n$.

\section{Conclusions}
\label{sec:5}

In this paper, the cell model was used to study
the behavior of fluid in a close vicinity of its critical
point. This is an interesting (because of fundamental
and applied aspects) and difficult (because
of a substantial role of fluctuation effects) issue for
analysis. The study of the relationship between the
density and the chemical potential at temperatures
$T<T_c$ made it possible to determine the corresponding
densities in the regions where the chemical potential
changes and obtain both the equation of state
and the binodal equation. Solutions of a certain cubic
equation were presented, which the equation of
state of the cell fluid model depends on. Their analysis
made it possible to describe the transition from
one solution to another when the chemical potential
tends to zero. On the basis of the obtained equation of
state, the pressure variation with the density growth
at various temperatures has been illustrated graphically.
Using the binodal equation, a binodal curve in a
narrow temperature interval was constructed for the
microscopic parameters of the Morse potential that
are inherent to sodium. As compared with the case
of the zero-mode approximation, the obtained dome
of the coexistence curve is wider and agrees better
with the results of computer simulation \cite{singh}.

\end{document}